\documentclass[
	letterpaper,% Paper size, use either a4paper or letterpaper
	10pt, % Default font size, can also use 11pt or 12pt, although this is not recommended
        unnumberedsections,
	twoside % Two side traditional mode where headers and footers change between odd and even pages, comment this option to make them fixed
]{styles}

\runninghead{Retrieval Augmented Generation Integrated Large Language Models in Smart Contract Vulnerability Detection} % A shortened article title to appear in the running head, leave this command empty for no running head

\footertext{\textit{}} % Text to appear in the footer, leave this command empty for no footer text

\title{Retrieval Augmented Generation\\Integrated Large Language Models in\\Smart Contract Vulnerability Detection} % Article title, use manual lines breaks (\\) to beautify the layout

\author{%
	Jeffy Yu\\
}

\date{\footnotesize Parallel Polis\\ San Francisco State University\\}

\renewcommand{\maketitlehookd}{%
	\begin{abstract}
		\noindent The rapid growth of Decentralized Finance (DeFi) has been accompanied by substantial financial losses due to smart contract vulnerabilities, underscoring the critical need for effective security auditing. With attacks becoming more frequent, the necessity and demand for auditing services has escalated. This especially creates a financial burden for independent developers and small businesses, who often have limited available funding for these services. Our study builds upon existing frameworks by integrating Retrieval-Augmented Generation (RAG) with large language models (LLMs), specifically employing GPT-4-1106 for its 128k token context window. We construct a vector store of 830 known vulnerable contracts, leveraging Pinecone for vector storage, OpenAI's text-embedding-ada-002 for embeddings, and LangChain to construct the RAG-LLM pipeline. Prompts were designed to provide a binary answer for vulnerability detection. We first test 52 smart contracts 40 times each against a provided vulnerability type, verifying the replicability and consistency of the RAG-LLM. Encouraging results were observed, with a 62.7\% success rate in guided detection of vulnerabilities. Second, we challenge the model under a "blind" audit setup, without the vulnerability type provided in the prompt, wherein 219 contracts undergo 40 tests each. This setup evaluates the general vulnerability detection capabilities without hinted context assistance. Under these conditions, a 60.71\% success rate was observed. While the results are promising, we still emphasize the need for human auditing at this time. We provide this study as a proof of concept for a cost-effective smart contract auditing process, moving towards democratic access to security. 
	\end{abstract}
}

\begin{document}

\maketitle % Output the title section

%----------------------------------------------------------------------------------------
%	ARTICLE CONTENTS
%----------------------------------------------------------------------------------------

\section{1. Introduction}

The rapid expansion of blockchain technology and decentralized finance (DeFi) has transformed the financial landscape, offering novel opportunities for peer-to-peer transactions and automated financial services. At the heart of this transformation are smart contracts—self-executing contracts that facilitate, verify, and enforce the terms of an agreement [1]. Despite their potential to streamline and secure transactions, smart contracts are not without risks. Vulnerabilities within these contracts can lead to significant financial losses and undermine trust in the DeFi ecosystem [12].

The critical need for effective smart contract auditing has never been more apparent. Traditional auditing methods, which rely on manual code reviews by security experts, are thorough but costly and time-intensive [2]. This poses a significant barrier for independent developers and small businesses that may lack the resources to afford such services. Consequently, there is a pressing demand for scalable, cost-effective solutions that can democratize access to smart contract security auditing, ensuring that all participants in the blockchain space can safeguard their assets and operations [15].

In response to this need, our study explores the integration of Retrieval-Augmented Generation (RAG) with large language models (LLMs) to enhance the detection of smart contract vulnerabilities. By leveraging the advanced capabilities of models such as GPT-4-1106, combined with a comprehensive vector store of known vulnerabilities, we aim to develop a robust auditing tool that is both accessible and reliable [3]. Our research evaluates the performance of this RAG-LLM system under both guided and blind conditions, providing insights into its potential and identifying areas for further refinement. Through this approach, we seek to contribute to the democratization of smart contract security, fostering a more inclusive and secure DeFi ecosystem.

%------------------------------------------------

\section{2. Background}

The explosive growth of decentralized finance (DeFi) has introduced new paradigms in financial transactions, leveraging blockchain technology to enable peer-to-peer lending, trading, and other financial services without traditional intermediaries. Central to these operations are smart contracts—self-executing contracts with the terms of the agreement directly written into code. While smart contracts offer numerous advantages, including automation, transparency, and reduced costs, they also pose significant security risks.

Smart contract vulnerabilities can lead to catastrophic financial losses, as evidenced by numerous high-profile attacks in the DeFi space. Common vulnerabilities include reentrancy attacks, integer overflows, and unchecked external calls [18]. These vulnerabilities exploit the logic and structure of smart contract code, often resulting in the unauthorized transfer of funds or other malicious outcomes.

To mitigate these risks, robust smart contract auditing is essential. Traditional auditing involves manual code reviews by security experts, which, although effective, are time-consuming and expensive [14, 16]. This creates a bottleneck for independent developers and small businesses, who may not have the resources to access high-quality auditing services.

The emergence of large language models (LLMs), such as OpenAI's GPT-4, has opened new avenues for automating parts of the auditing process. LLMs have demonstrated remarkable capabilities in understanding and generating human-like text, making them suitable for tasks that involve code analysis and pattern recognition. However, LLMs on their own are limited by their training data and may lack the specificity required for thorough smart contract auditing.

To address these limitations, the concept of Retrieval-Augmented Generation (RAG) has been introduced. RAG combines the generative abilities of LLMs with retrieval mechanisms that access a curated database of relevant information. This hybrid approach enables the model to generate more accurate and contextually relevant responses by referencing external sources of knowledge.

In the context of smart contract auditing, a RAG-LLM system can retrieve examples of known vulnerabilities from a vector store, enhancing its ability to identify similar issues in new contracts. By integrating this technology, we aim to develop a scalable and cost-effective solution that democratizes access to smart contract security auditing.

Our study builds upon this framework, employing GPT-4-1106 with a 128k token context window, Pinecone for vector storage, and OpenAI’s text-embedding-ada-002 for generating embeddings. We construct a vector store of 830 known vulnerable contracts and design prompts to test the model’s ability to detect vulnerabilities under both guided and blind conditions.

The experimental results from our study show promising success rates in vulnerability detection, demonstrating the potential of RAG-LLMs to provide reliable and accessible smart contract auditing services. However, challenges remain, particularly in ensuring the unbiased evaluation of model performance and addressing the variability in detection accuracy across different types of vulnerabilities.

%------------------------------------------------
\section{3. Democratic Access}

The advent of blockchain technology and decentralized finance (DeFi) has revolutionized the way financial transactions are conducted, offering unprecedented transparency, security, and decentralization. However, the increasing complexity and proliferation of smart contracts have introduced significant security challenges [13]. Vulnerabilities in smart contracts can lead to substantial financial losses, undermining trust in these systems and hindering widespread adoption [17]. This underscores the critical need for accessible and reliable smart contract auditing services.

Traditional smart contract auditing is typically conducted by specialized firms that offer thorough but expensive services. This creates a significant barrier for independent developers and small businesses that may lack the financial resources to afford these audits. Consequently, the democratization of smart contract security auditing is essential to ensure equitable access to these vital services, fostering a more inclusive and secure DeFi ecosystem.

Democratic access to security auditing means making robust and reliable auditing tools available to all participants in the blockchain space, regardless of their size or resources. By leveraging advanced technologies such as Retrieval-Augmented Generation (RAG) integrated with large language models (LLMs), our project aims to provide a scalable, cost-effective solution that democratizes access to smart contract security.

RAG-LLMs enhance the auditing process by combining the generative capabilities of LLMs with the precision of retrieval mechanisms, allowing for the identification and analysis of vulnerabilities in smart contracts. This integration not only improves the accuracy of vulnerability detection but also makes these tools more accessible to a broader audience. Independent developers and small enterprises can leverage these advanced auditing capabilities without the prohibitive costs associated with traditional auditing services.

The motivation behind this project is to empower all stakeholders in the blockchain ecosystem with the tools necessary to ensure the security of their smart contracts. By providing decentralized, trustless auditing capabilities, we aim to foster greater trust in blockchain technologies. This trust is crucial for the widespread adoption and integration of blockchain into various sectors, from finance to supply chain management.

In a trustless environment, where transactions and operations are executed without the need for intermediaries, the integrity and security of smart contracts become paramount [19]. Democratic access to auditing tools ensures that all participants can independently verify the security of their smart contracts, promoting a more resilient and trustworthy blockchain infrastructure.

Moreover, as the DeFi space continues to evolve, the scalability of our approach allows it to keep pace with the growing number of smart contracts and their increasing complexity. By continually enhancing the capabilities of RAG-LLMs and making these tools widely available, we support the sustainable growth of a secure and inclusive DeFi ecosystem.

%------------------------------------------------

\section{4. RAG-LLM Pipeline}

\begin{figure}[ht]
	\centering
        \includegraphics[width=0.5\textwidth]{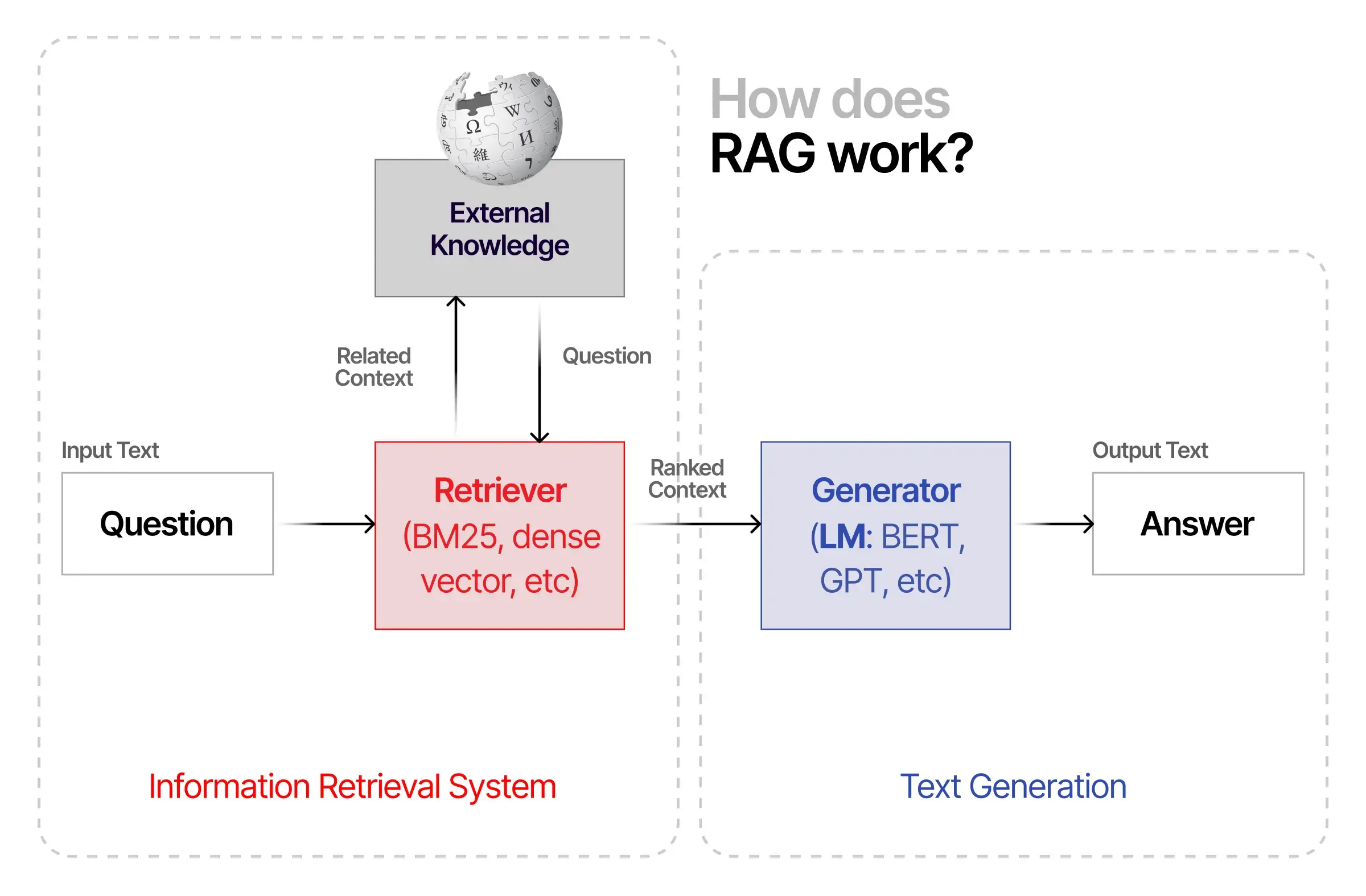}
	\captionsetup{justification=raggedright,singlelinecheck=false}\caption{RAG LLM Pipeline. Adapted from "How to Connect LLM to External Sources Using RAG?" by Rajeev Sharma, \href{https://markovate.com/blog/connect-llm-using-rag/}{https://markovate.com/blog/connect-llm-using-rag/}.}
	\label{fig:tcanther}
\end{figure}

The integration of Retrieval-Augmented Generation (RAG) with large language models (LLMs) extends their capacity to generate responses by incorporating external knowledge sources. This paradigm enhances model performance on queries involving information beyond its initial training data. RAG operates by enriching the model's context with relevant content from a vector store, expanding the data reservoir accessible during inference. This dual retrieval mechanism distinguishes RAG-LLMs from traditional LLMs, elevating their ability to synthesize and contextualize responses by concurrently utilizing the model's internal knowledge and external databases.

Our system capitalizes on integrating RAG with LLMs to apply their combined strengths for the domain of smart contract security. We employ GPT-4-1106 due to its extensive 128k token context window that enables the processing of lengthy prompts combined with information retrieval context [4]. The system is setup with a vector store containing embeddings for 830 known vulnerable smart contracts. We utilize Pinecone for vector indexing and storage due to its efficient scaling and retrieval capabilities, coupled with OpenAI's text-embedding-ada-002 to generate the embeddings [4,5]. The OpenAI text-embedding-ada-002 model is capable of generating embeddings for mixed content, encompassing both natural language and code, which is ideal for our dataset that includes Solidity smart contracts interlaced with natural language comments [4]. LangChain, a framework optimized for constructing RAG-LLM pipelines, facilitates our retrieval-generative processes [6].

The system is designed to enhance the quality of responses by retrieving pertinent-vulnerability patterns from the vector store of known vulnerable smart contracts. Upon receiving a user prompt potentially indicative of insecure coding practices, our setup identifies similarities within the vector store. A match suggests parallels between the user's code and pre-identified vulnerabilities, flagging a likely security issue.

\subsection{4.1 Vulnerable Contract Dataset}

In order to construct an expansive and robust dataset for our RAG-LLM pipeline, we collected a total of 830 vulnerable smart contracts. These contracts were sourced from public Github repositories known to house collections of compromised or flawed smart contract code. The repositories utilized include "(Not So) Smart Contracts" by Trail of Bits, "DeFiHackLabs" and "DeFiVulnLabs" by SunWeb3Sec, and "Smart Contract VulnDB" by tintinweb [7,8,9].

The dataset comprises a wide spectrum of vulnerability types which are essential to develop a comprehensive understanding of the security pitfalls within smart contracts. Some of the vulnerabilities included are:

\begin{itemize}
    \item Bad randomness
    \item Denial of service
    \item Forced ether reception
    \item Honeypots
    \item Incorrect interface implementation
    \item Integer overflow
    \item Race condition occurrences
    \item Reentrancy attacks
    \item Unchecked external calls
    \item Unprotected functions
    \item Variable shadowing
    \item Misnamed constructors
\end{itemize}

\noindent This list is not exhaustive. \\

This variation in the dataset aims to ensure that the system is well-equipped to recognize and provide context for a diverse array of security weaknesses.

The selection criteria for contract inclusion was dictated by the requirement to have a diverse set of vulnerabilities rather than by volume. While a greater number of contracts per vulnerability type may have contributed to a more substantial dataset, the prevalence of vulnerable contract datasets is limited, offering a pragmatic cap to the dataset size. 

Moreover, we reserved a significant portion of gathered contracts exclusively for validation purposes to avoid direct data overlap between the RAG vectorstore and subsequent testing. While we had to sacrifice data for vectorstore creation, this measure is taken to ensure the integrity of our experimental evaluation, preventing the model from simply regurgitating exact code matches without genuine understanding.

\subsection{4.2 Data Processing and Embedding}

We initiated the data processing by loading .sol files—designating them as unstructured text documents. We employed the LangChain DirectoryLoader to retrieve Solidity smart contracts from a specified directory. Following loading, the documents underwent fragmentation into smaller text chunks. To accommodate the maximum context window of our selected model, each chunk was composed of 1024 tokens, a size that was large enough to prevent truncation of functions while reasonably fitting in the context window of the GPT-4 model.

LangChain's TokenTextSplitter, utilizing the tokenisation method from TikToken, was selected as the tool to divide the contracts into token-sized chunks. TikToken provided two key benefits for processing our dataset:

\begin{itemize}
    \item Reversibility – The tokenization process is designed to be reversible, ensuring that the transformed tokens can be converted back into the original text without any loss of information, which is imperative for the integrity and fidelity of the dataset.
    \item Generalization Capability – TikToken is engineered to generalize beyond its training corpus, enabling it to tokenize unfamiliar text sequences effectively. This characteristic is particularly beneficial for parsing Solidity code, as its syntax may deviate considerably from the data TikToken was originally trained on.
\end{itemize}

Subsequently, embeddings were created for the tokenized chunks using the OpenAI API with the "text-embedding-ada-002" model. This model was specifically selected due to its proficiency in encoding a blend of natural language and programming code, which makes it particularly suited to the mixed content of smart contracts that incorporates both Solidity code and natural language code annotations.

For the vector store, we configured Pinecone with a 'p2' index option, which was chosen for its rapid search capabilities. Upon passing the embeddings to Pinecone, we created a vector database with 939 vectors, each with the embeddings (of 1536 dimensions) and corresponding metadata for each code chunk from the smart contracts.

\subsection{4.3 RAG with LangChain}

Our system incorporates Retrieval-Augmented Generation (RAG) technology within the LangChain framework to analyze smart contracts for potential vulnerabilities. LangChain is a versatile toolkit that facilitates the creation and deployment of chains—a series of operations involving language models and other processing steps.

Our LLM, specifically ChatOpenAI configured with the GPT-4 model "gpt-4-1106-preview," operates with set parameters such as a temperature of 0.7 for response variability and a maximum token allowance. We chose this model due to the expanded 128,000 token window, which allows fitting the full user smart contract along with all retrieved code from vector search in a single prompt. For our experiment, we set a maximum token response of 5, as our prompts instructed a binary YES or NO response. When using this system for open ended analysis and identification of potential fixes, a larger maximum token response can be set up to 4095 tokens.

We engage Pinecone's services to retrieve relevant code from our established index, populated with our embeddings. This index functions as the RAG vector store from which our retriever conducts searches. The searcher is set to return the top 'k' results, in this case. 

We elected to return the top 5 most relevant documents for a given query. We chose this based on the reasoning that having a larger or unbounded quantity may dilute the prompt due to the sheer amount of data returned. Additionally, we believe the focus on the prompt would be diluted, on both the content for in-context learning and the user prompt, if too much context is provided. Therefore, we limit retrieval to 5 similar documents.

The core of our RAG system is powered by a prompt template which scripts the interaction between the user’s input and the LLM for auditing smart contracts. This template guides the LLM to focus solely on identifying any vulnerabilities that may exist within the user’s question, which comprises the source code of a smart contract. The template disallows the analysis of vulnerabilities detailed in the retrieved contextual information, ensuring that the LLM's response adheres strictly to the information provided in the user’s query.

%------------------------------------------------

\section{5. Experiment Design}

Our experiment design is aligned with evaluating the efficacy of the RAG-LLM system in identifying smart contract vulnerabilities. The testing methodology bifurcates into two distinct phases: guided detection and blind detection. Guided detection involves an explicit indication of the vulnerability type and a corresponding description provided within the prompt.

Blind detection, conversely, relies solely on the RAG retrieved context and the smart contract code in question, without explicit prompt instructions regarding the vulnerability type.

\subsection{5.1 Guided vs. Blind Audit}

The dichotomy in our experiment design, specifically the demarcation of guided versus blind audit phases, is structured to test the RAG-LLM system's capabilities under contrasting conditions.

In the guided audit phase, the prompts are engineered to include explicit mention of the vulnerability type accompanied by its description. This phase evaluates the system's precision when the language model is provided with a clear framework of the vulnerability expected to be found within the smart contract. The benefits of this method include testing the LLM's comprehension of specific vulnerability descriptors and its ability to correlate those with patterns detected in the code. Moreover, we assess whether the presence of guided information aids or biases the decision-making accuracy of the model. The reasoning behind incorporating a guided audit is to measure the added value or possible detraction of closely directed context in enabling the model to pinpoint exact vulnerabilities.

In contrast, the blind audit phase assesses the language model's unsupervised analytical proficiency. It eliminates any direct guidance toward a particular vulnerability, instead relying solely on the system's RAG retrieval mechanism to provide context. This method tests the model's ability to generalize from the vectorized corpus of smart contract vulnerabilities and its intrinsic understanding derived from initial training. It pushes the boundaries of the system's autonomous functionality by requiring it to discern and identify vulnerabilities purely based on the contextual relevance of the presented code. This approach also provides insights into potential model limitations and uncovers areas where in-context learning might need to be refined for accurate unsupervised decision-making.

Both phases are critical for a comprehensive analysis of the system design. The guided audits act as a controlled environment to monitor the system's performance with clear indicators, similar to assessing its capability to perform targeted code analysis. The blind audits test the system's generalization, replication of logical reasoning, and its independent inferential abilities, mirroring a more practical real-world scenario where explicit guidance is not always available.

\subsection{5.2 Testing Dataset Curation}

The curation of our testing dataset was methodically approached to ensure rigorous assessment in both phases of our experiment. For the initial phase focused on guided detection, we utilized the set of 52 smart contracts catalogued in Table 1 of the study by David et al [11, Table 1]. This particular collection was initially sampled and presented by Zhou et al. in their work on decentralized finance (DeFi) attacks [12]. We have replicated this enumerated dataset in Table 1 of our paper, and the vulnerability types along with their descriptions were directly sourced from Table 6 in David et al.'s publication [number, Table. 6]. Adopting this dataset enables us to undertake a level of replication from David et al.'s experimental design while adapting the study to evaluate our RAG system's specific parameters.

For the second phase of blind detection, our experiment expands the evaluation scope to include the broadly composed dataset of 219 smart contracts, as detailed in the comprehensive "Systematization of Knowledge" (SoK) by Zhou et al. The use of this extensive dataset allows for an exploration into the generative capabilities of our RAG system across a diverse and realistic range of smart contracts, derived from actual DeFi incident analyses. This phase is critical in assessing the model's performance in a more autonomous scenario without prior indication of potential vulnerabilities.

\subsection{5.3 Prompt Engineering}

Prompt engineering plays a critical role in the effectiveness of language model interactions, particularly when specialized tasks such as smart contract audits are involved. In our methodology, we designed the prompts to elicit specific responses from the model regarding potential vulnerabilities in smart contract code.

In the initial stages of our study, we considered replicating the experimental design and prompt engineering methods employed by David et al., with the intention of conducting a comparative analysis. David et al. conducted tests on 52 smart contracts, each against 38 different vulnerability types; a process that involved evaluating a smart contract with a known vulnerability using a corresponding prompt explaining that vulnerability, and then separately using prompts for the 37 other vulnerability types to assess the language model's susceptibility to false positives and negatives. 

However, upon further consideration, we determined that our experimental setup, which inherently includes a RAG component that retrieves contextually relevant code snippets, was not conducive to the same methodology. In their framework, each vulnerability type is evaluated in isolation to test the model's identification accuracy. In contrast, our RAG system is designed to pull examples of code similar to the user input — implying that, for a contract known to be susceptible to reentrancy attacks, the retrieved code would primarily illustrate reentrancy vulnerabilities.

This retrieval specificity introduces a complication; when the RAG component fetches documents that align with the primary vulnerability, the in-context learning becomes biased towards that vulnerability type. As a result, evaluating the contract against other vulnerability categories, such as integer overflow, might lead to misleading outcomes since the retrieved context would not pertain to the vulnerability in question. Therefore, our RAG-enhanced setup naturally skews the language model towards a correct identification of the dominant vulnerability, thereby undermining the validity of a testing approach designed around the identification of different potential vulnerabilities.

Consequently, we elected not to directly replicate David et al.'s experimental framework, but rather to retain certain elements from their study, including the use of analogous prompts and datasets. This strategic adjustment allows us to test for the replicability and consistency of our RAG-enhanced system in identifying vulnerabilities within smart contracts. We aim to measure the repeatable efficiency of our RAG integration and gain insights into the variability and potential non-determinism within large language model responses. 

This approach not only preserves the integrity of our experimental setup, but also contributes to the assessment of our system's precision and the exploration of the intrinsic randomness that may affect the outcomes of language model-based analyses.

The prompt for phase one is shown in Figure 2.

\begin{figure}[ht]
\begin{lstlisting}[language={},frame=tlrb]
You are an AI Smart Contract auditor agent in a RAG system. 
We have performed a vector search of known smart contract 
vulnerabilities based on the code in the USER QUESTION.
The results are below:

RELEVANT_VULNERABILITIES: {context}

With this knowledge, review the following smart contract code 
in USER QUESTION in detail and very thoroughly.
ONLY indentify vulnerabilities in the USER QUESTION, do not 
analyze the RELEVANT_VULNERABILITIES.

Think step by step, carefully. 
Is the following smart contract vulnerable to '{vulnerability_type}' attacks? 
Reply with YES or NO only. Do not be verbose. 
Think carefully but only answer with YES or NO! To help you, 
find here a definition of a '{vulnerability_type}' attack: {vulnerability_description}

USER QUESTION: {question}
\end{lstlisting}
\caption{Phase one prompt template used for RAG with LangChain}
\label{fig:prompt_template}
\end{figure}

The placeholders within the prompt are replaced with specific information for each iteration of the model's use:

\begin{enumerate}
    \item \texttt{\{context\}} - Content retrieved from the vector store (Pinecone), which is semantically relevant to the user's question, replaces this placeholder. These documents typically hold patterns, code snippets, or commentaries pertaining to recognized vulnerabilities in smart contracts.
    \item \texttt{\{question\}} - Into this placeholder, the actual code snippet or smart contract furnished by the user for analysis is inserted. It represents the text that the model is tasked with examining to ascertain the existence or non-existence of vulnerabilities.
    \item \texttt{\{vulnerability\_description\}}, \texttt{\{vulnerability\_type\}} - These placeholders are filled with the specific type and definition of a vulnerability from a pre-curated list. This specific information directs the model to assess the query with regard to a singular vulnerability type, testing guided detection.
\end{enumerate}

The design choice to solicit binary responses—'YES' or 'NO'—from the language model is underpinned by methodological considerations that enhance the reliability and objectivity of the analysis. By constraining the language model's output in this manner, the need for human interpretation of more verbose responses is eliminated, thereby reducing the potential introduction of subjective bias into the evaluation process. A binary decision format ensures deterministic results that are straightforward to classify, categorize, and analyze statistically.

%------------------------------------------------

\section{6. Phase One}

Presented here are the outcomes of the first phase of our experimentation, which sought to evaluate the precision of our RAG-LLM system in a guided setting. The results derived from this phase provide insight into the system's capability to correctly identify specific vulnerabilities when provided with contextual cues in the form of vulnerability types and descriptions.

\subsection{6.1 Results}

The experiment involved the application of the RAG-LLM system to 52 smart contracts, each evaluated against a multitude of vulnerability types explicitly mentioned within the prompts. The tabulated data in Table \ref{tab:phaseonepercents} compiles the aggregated results from these individual assessments.

As depicted, the system achieved a total of 1303 successful identifications, representing a success rate of approximately 62.7\%. Conversely, the number of instances where the system failed to correctly identify a vulnerability—or incorrectly identified the presence of one that was not present—stood at 777, accounting for 37.3\% of the evaluations. The overall success rate of 0.6270 attests to the system's capability to correctly identify vulnerabilities with modest reliability when provided with specific, indicative context.

\begin{table}[ht]
	\caption{Phase One Results, totals from 52 smart contracts tested}
	\centering
	\begin{tabular}{l r r}
		\toprule
		\textbf{Overall Result} & \textbf{Count} & \textbf{Percentage}\\
		\midrule
		Successes & 1303 & 0.627 \\
		Failures & 777 & 0.373 \\
		\bottomrule
	\end{tabular}
	\label{tab:phaseonepercents}
\end{table}

\begin{figure*}[tp]
\centering
\begin{tikzpicture}
\begin{axis}[
    width=\textwidth, % specify the width of the graph
    height=9cm, % you can adjust the height as needed
    bar width=0.007\textwidth,
    ybar,
    ymin=0, ymax=110,
    ylabel={Success Rate (\%)},
    enlarge x limits=0.02, % add space between the bars
    symbolic x coords={
        1., 2., 3., 4., 5., 6., 7., 8., 9., 10.,
        11., 12., 13., 14., 15., 16., 17., 18., 19., 20.,
        21., 22., 23., 24., 25., 26., 27., 28., 29., 30.,
        31., 32., 33., 34., 35., 36., 37., 38., 39., 40.,
        41., 42., 43., 44., 45., 46., 47., 48., 49., 50.,
        51., 52.
    },
    xtick=data,
    x tick label style={
        rotate=90, 
        anchor=east
    },
    ylabel style={font=\small},
    xlabel style={font=\small},
    ticklabel style={font=\small},
    nodes near coords,
    nodes near coords align={vertical},
    nodes near coords style={font=\tiny}
]

\addplot coordinates {
    (1., 77.5)
    (2., 97.5)
    (3., 25.0)
    (4., 100)
    (5., 95.0)
    (6., 75.0)
    (7., 57.5)
    (8., 7.5)
    (9., 52.5)
    (10., 25.0)
    (11., 100)
    (12., 0.0)
    (13., 75.0)
    (14., 65.0)
    (15., 15.0)
    (16., 2.5)
    (17., 60.0)
    (18., 100)
    (19., 95.0)
    (20., 100)
    (21., 55.0)
    (22., 70.0)
    (23., 15.0)
    (24., 82.5)
    (25., 100)
    (26., 15.0)
    (27., 60.526)
    (28., 97.5)
    (29., 87.5)
    (30., 82.5)
    (31., 70.0)
    (32., 12.5)
    (33., 95.0)
    (34., 5.0)
    (35., 85.0)
    (36., 52.5)
    (37., 72.5)
    (38., 17.5)
    (39., 12.5)
    (40., 100)
    (41., 45.0)
    (42., 100)
    (43., 7.5)
    (44., 82.5)
    (45., 67.5)
    (46., 92.5)
    (47., 85.0)
    (48., 90.0)
    (49., 77.5)
    (50., 92.5)
    (51., 52.5)
    (52., 55.0)
};

\end{axis}
\end{tikzpicture}
\caption{Phase One Results, individual smart contract efficacy. See Appendix Table 3 for corresponding smart contract.}
\label{fig:prompt_template}
\end{figure*}
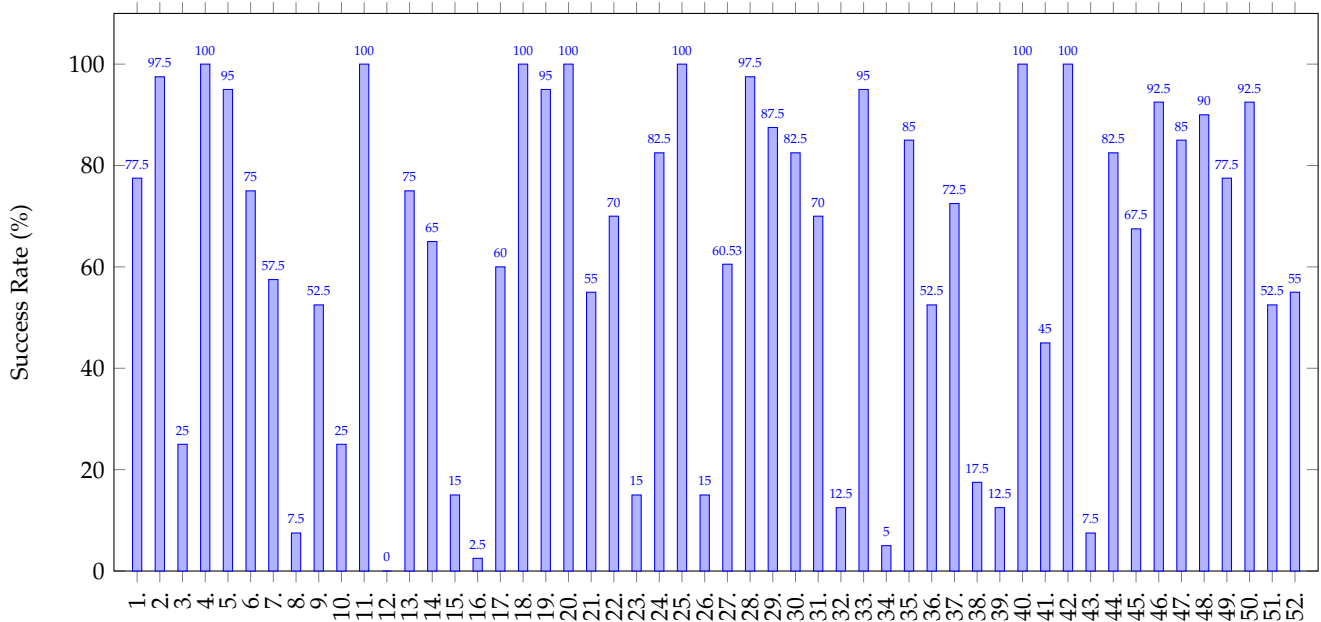

\subsection{6.2 Discussion}

In comparing our findings with those presented by David et al., we observe an improvement in the success rate of vulnerability detection. Their study, employing a GPT-4 model with a 32k token window, achieved a success rate of approximately 59.76\%, calculated by aggregating 32 true positives (TP) and 1128 true negatives (TN) over the total number of data points (1950), with the failure rate registering around 40.24\% derived from 740 false positives (FP) and 41 false negatives (FN). Our usage of GPT-4-1106-preview yielded a slightly higher success rate of 62.7\% and a correspondingly lower failure rate of 37.3\%, likely attributed to the expanded token windows and retrieved context. We note that these results may not be directly comparable due to the differences in experiment design.

The dataset analyzed revealed no discernible pattern in the performance of the RAG-LLM system across different vulnerability types. Variability in the success rates was evident, with specific contracts having the same vulnerability type demonstrating significant deviations in how effectively they were audited. This lack of consistency suggests that the model's performance is influenced by more nuanced factors than simply the category of vulnerability being scrutinized.

Moreover, the results demonstrated the absence of uniformity in classification accuracy, as very few smart contracts were consistently identified with 0\% or 100\% efficacy. This variability implies an element of randomness in the model's classification process, whereby the same input does not guarantee identical results across multiple audits. To construct a metric of classification reliability, we tested each smart contract 40 times against a single specified vulnerability. The derived percentages display the inherent volatility in the model's decision-making, providing quantifiable insights into the repeatability of results. The practical implication of this finding suggests that multiple runs may be needed to corroborate the model's conclusions reliably when utilizing the RAG-LLM system for smart contract audits.

In the retrieval process, the vectorstore was queried using the code provided from smart contract being tested without incorporating the vulnerability type or description. Our system only fetched results relying solely on the semantic similarity inherent in the smart contract code and the vectorstore. Future iterations of testing could explore "guided retrieval," incorporating the vulnerability type and description into the retrieval query, to scrutinize the impact of this information on the retrieval process and performance of vulnerability identification.

It is crucial to acknowledge a limitation in our current design; the binary YES/NO response mechanism does not differentiate between false positives and false negatives. As such, any positive (YES) identification cannot be conclusively verified within the scope of our analysis due to the absence of an underlying ground truth for validation. This binary outcome design focused on simplicity and elimination of subjective biases implies that while we can quantify success and failure rates, we lack granular insight into the nature of inaccuracies that arise from the model's processing. Future studies may address this by incorporating a mechanism to assess responses against a verifiable ground truth, enabling distinction between false positives and negatives which would enhance the depth of system evaluation.

%------------------------------------------------

\section{7. Phase Two}

Building on the methodologies established in the guided audit phase, Phase Two introduces a more stringent testing environment for evaluating the Retrieval-Augmented Generation (RAG) system's capabilities in detecting smart contract vulnerabilities. In this phase, the prompts provided to the language model exclude explicit vulnerability types, challenging the system to rely solely on its retrieval mechanism and intrinsic understanding of smart contract security.

The blind audit phase is designed to test the model's unsupervised analytical proficiency. By removing direct guidance towards specific vulnerabilities, this phase emphasizes the model's ability to generalize from the vectorized corpus of known vulnerabilities and apply this knowledge autonomously.

\subsection{7.1 Results}

For brevity, the full results for phase two are listed in the appendix, within Table 4.

In Phase Two, the RAG-LLM system was tested under a blind audit setup, evaluating its ability to detect vulnerabilities without explicit prompt instructions regarding the vulnerability type. The experiment involved 219 smart contracts, each undergoing 40 tests. The aggregated results are summarized in Table 2, showing a success rate of approximately 60.71\%.

\begin{table}[ht]
	\caption{Phase Two Results, totals from 220 smart contracts tested}
	\centering
	\begin{tabular}{l r r}
		\toprule
		\textbf{Overall Result} & \textbf{Count} & \textbf{Percentage}\\
		\midrule
		Successes & 1303 & 0.627 \\
		Failures & 777 & 0.373 \\
		\bottomrule
	\end{tabular}
	\label{tab:phaseonepercents}
\end{table}

The system's success rate in this phase, although slightly lower than the guided detection phase, demonstrates its capability to identify vulnerabilities with reasonable accuracy even without explicit guidance

\subsection{7.2 Discussion}

The results from Phase Two indicate that the RAG-LLM system maintains a robust performance in a more challenging blind audit scenario. The overall success rate of 60.71\% is encouraging, showcasing the model's potential to generalize and detect vulnerabilities autonomously.

\subsubsection{Comparison To Phase One}
The guided detection phase yielded a higher success rate of 62.7\%, benefiting from explicit vulnerability type cues.

The slight drop in success rate during the blind audit phase (60.71\%) suggests that the model's performance is influenced by the absence of direct guidance, highlighting areas for improvement in its unsupervised analytical capabilities.

\subsubsection{Variability}
The performance variability observed across different smart contracts and vulnerability types indicates that certain vulnerabilities are more challenging for the model to detect without explicit context. For instance, smart contracts with reentrancy vulnerabilities showed a significant deviation in detection accuracy, with success rates ranging from 0\% to 97.5\%.

This variability underscores the model's dependency on the quality and relevance of retrieved contextual information. It suggests that while the model can generalize, its effectiveness is enhanced when relevant examples are retrieved and presented.  For instance:

\subsubsection{On-chain Oracle Manipulation:} Success rates varied significantly, with some contracts achieving near-perfect detection accuracy (e.g., 97.5\% for contract 172) and others much lower (e.g., 30\% for contract 174).

\subsubsection{Absence of Code Logic or Sanity Check:} This vulnerability type also showed a wide range of success rates, from as low as 0\% (contract 7) to as high as 100\% (contracts 109 and 110).

\subsubsection{Reentrancy:} Detection accuracy for reentrancy vulnerabilities ranged from 0\% for some contracts (e.g., contract 47) to 97.5\% for others (e.g., contract 14), indicating inconsistency in the model's ability to detect this vulnerability type without explicit guidance.

\subsubsection{Backdoor / Honeypot:} This category showed relatively high detection rates, with most contracts achieving success rates above 70\%, suggesting that the model is generally proficient in identifying these vulnerabilities even in a blind audit scenario.

The data also highlights specific contracts where the model struggled, such as those involving "Deployment Mistakes" and "Function Visibility Errors," which showed lower and more variable success rates. These findings suggest that the model's retrieval mechanism and its contextual understanding are critical factors influencing its performance. Contracts like those with "Unfair Slippage Protection" and "Improper Asset Locks" had particularly low success rates, pointing to areas where the model's detection capabilities could be enhanced with better contextual information and retrieval techniques .

In conclusion, while the RAG-LLM system demonstrates substantial promise in autonomous vulnerability detection, the results from Phase Two highlight the importance of context and guidance in maximizing its effectiveness. Continued enhancements in retrieval strategies and model training will be critical in advancing the system's capabilities for practical deployment in smart contract security auditing.

%------------------------------------------------

\section{8. Considerations}
In reflecting on the comprehensive scope and findings of this project, several key considerations emerge that warrant attention for future research and practical application in the field of smart contract auditing. These considerations encompass limitations, potential improvements, and broader implications of our study.

\subsection{8.1 Training Data Integrity}
The language models (LLMs) utilized in this study, such as GPT-4-1106, operate as black boxes with undisclosed training datasets. Consequently, there exists a possibility that these models have been exposed to some of the vulnerabilities and smart contract codes we tested. This exposure could introduce biases, potentially inflating the models' success rates. Future work should aim to develop and test models on completely novel datasets to ensure unbiased evaluation of their capabilities.

\subsection{8.2 Binary Classification Compliance}
One observed limitation is the occasional non-compliance of LLMs with binary classification prompts. Instances where the models provided verbose explanations rather than simple "YES" or "NO" responses necessitated manual verification to maintain metric coherence. Future research should explore refining prompt engineering techniques or model adjustments to enhance compliance with binary classification tasks.

\subsection{8.3 False Positives and Ground Truth Ambiguity}
The ground truth for the set of 52 vulnerable DeFi contracts used in our study is not definitive. These contracts might harbor additional, unreported vulnerabilities, which could affect the accuracy of our false positive metrics. Future studies should incorporate a mechanism for more thorough verification of contract vulnerabilities, possibly involving expert human auditors or cross-referencing multiple vulnerability databases.

\subsection{8.4 Context Limitations and Truncation Strategies}
We encountered context length limitations with the GPT-4-1106 model, particularly with lengthy smart contract codes. Our approach involved naive truncation to fit the model's context window, which may have resulted in loss of critical information. Future work should investigate more sophisticated methods, such as dynamic context compression or multi-pass querying, to handle extensive source codes without truncation.

\subsection{8.5 Impact of Extended Context on Model Performance}
Qualitative observations suggest that providing longer contexts might reduce the likelihood of models adhering to binary classification tasks. Understanding the internal mechanisms of LLMs, especially those capable of handling extended contexts, could shed light on this phenomenon. Future research should quantify the impact of longer contexts on model performance and classification accuracy, potentially leading to improved prompt structuring.

\subsection{8.6 Security Implications and Ethical Considerations}
While our study demonstrates the potential of RAG-LLMs in smart contract auditing, it also raises security and ethical considerations. The reliance on automated systems for vulnerability detection could inadvertently introduce new risks if these systems are not rigorously validated. Ethical considerations regarding the transparency and accountability of AI-driven audits should be addressed, ensuring that human oversight remains integral to the auditing process.

\subsection{8.7 Scalability and Real-World Application}
The scalability of our RAG-LLM approach for large-scale, real-world applications presents another critical consideration. While our experiments show promising results in controlled settings, deploying this system in diverse, dynamic environments such as decentralized finance ecosystems will require extensive testing and optimization. Future research should focus on scalability, robustness, and integration with existing security frameworks.

In summary, while our study marks significant progress in the field of smart contract auditing, these considerations highlight areas for further research and development. Addressing these aspects will be crucial in advancing the efficacy, reliability, and ethical deployment of AI-driven vulnerability detection systems.

%------------------------------------------------

\section{9. Conclusion}

The rapid evolution of Decentralized Finance (DeFi) necessitates robust and scalable security auditing mechanisms to mitigate the risks posed by smart contract vulnerabilities. Our study explores the integration of Retrieval-Augmented Generation (RAG) with large language models (LLMs) to enhance the detection of these vulnerabilities, marking a significant step towards democratizing access to smart contract security.

Our approach leverages the extensive context windows of GPT-4-1106, combined with a meticulously curated dataset of known vulnerabilities, to construct a powerful RAG-LLM pipeline. Through a two-phased experimental design, we evaluated the system's performance under guided and blind audit conditions, yielding promising results that underscore the potential of RAG-LLMs in this domain.

In the guided audit phase, the explicit provision of vulnerability types facilitated a higher success rate of 62.7\%, demonstrating the system's ability to effectively utilize contextual cues. Conversely, the blind audit phase, which challenged the model to detect vulnerabilities without explicit guidance, resulted in a slightly lower success rate of 60.71\%. This indicates the model's robust generalization capabilities, albeit with room for improvement in unsupervised contexts.

The variability observed in detection accuracy across different smart contracts and vulnerability types highlights the nuanced nature of smart contract auditing. Certain vulnerabilities, such as reentrancy, proved more challenging to detect consistently, pointing to the critical role of contextually relevant retrieval in enhancing model performance. These findings suggest that while the RAG-LLM system is a powerful tool, its efficacy can be significantly influenced by the quality of the retrieved contextual information.

Despite the promising results, our study also reveals several limitations. The potential biases introduced by the unknown training datasets of the LLMs, the occasional non-compliance with binary classification prompts, and the challenges posed by context length limitations all underscore the need for further refinement and validation. Moreover, the ethical implications of relying on automated systems for vulnerability detection warrant careful consideration, ensuring that human oversight remains a cornerstone of the auditing process.

Overall, our study demonstrates the feasibility and potential of using RAG-LLMs for smart contract vulnerability detection, offering a cost-effective and scalable solution that could significantly enhance the security landscape of DeFi ecosystems. However, realizing the full potential of this approach will require addressing the identified limitations and exploring new avenues for improvement.

\subsection{9.1 Future Work}

Future research should build on our findings, addressing the following key areas:

\begin{itemize}
    \item \textbf{Training Data Integrity:} Develop and evaluate models on novel, unbiased datasets to mitigate potential training data biases.
    \item \textbf{Binary Classification Compliance:} Refine prompt engineering techniques and model adjustments to ensure consistent compliance with binary classification tasks.
    \item \textbf{False Positive Verification:} Incorporate mechanisms for thorough verification of smart contract vulnerabilities, involving expert human auditors or cross-referencing multiple databases.
    \item \textbf{Context Handling:} Investigate sophisticated methods for handling lengthy smart contract codes, such as dynamic context compression or multi-pass querying.
    \item \textbf{Impact of Extended Contexts:} Quantify the impact of extended contexts on model performance and classification accuracy to inform improved prompt structuring.
    \item \textbf{Ethical Considerations:} Address ethical issues related to the transparency and accountability of AI-driven audits, ensuring robust human oversight.
    \item \textbf{Scalability:} Focus on the scalability and robustness of the RAG-LLM system for deployment in diverse, real-world DeFi environments.
    \item \textbf{Integration with Security Frameworks:} Explore the integration of RAG-LLM systems with existing security frameworks to enhance overall effectiveness.
    \item \textbf{Long-term Performance Monitoring:} Establish mechanisms for continuous monitoring and evaluation of the system’s performance in real-world applications to ensure sustained efficacy.
\end{itemize}

By addressing these areas, future research can further advance the capabilities of RAG-LLMs, making them an indispensable tool in the ongoing effort to secure the rapidly evolving landscape of decentralized finance.

%------------------------------------------------

%----------------------------------------------------------------------------------------
%	 REFERENCES
%----------------------------------------------------------------------------------------

%----------------------------------------------------------------------------------------

\onecolumn

\section{Appendix}
\begin{longtable}{c l l}
	\caption{Phase one smart contracts, sourced from David et al, Table 6 [10] and DeFi SoK [11]} \\
	\toprule
	ID & Address & Description\\
	\midrule
	\endfirsthead % Everything above this will be repeated on every page's header
	\caption{Phase one smart contracts (continued)} \\
	\toprule
	ID & Address & Description\\
	\midrule
	\endhead % Everything above this will be placed at the top of the continuation pages
	\bottomrule
	\endfoot % Footer for all pages except the last, usually empty
	\bottomrule
	\endlastfoot % Footer for the last page of the table
1 & 0xe952...1659 & Reentrancy, Token standard incompatibility \\
2 & 0x55db...3830 & On-chain oracle manipulation, Absence of code logic or sanity check \\
3 & 0x4e3f...9d2b & Governance attack \\
4 & 0x833e...743d & On-chain oracle manipulation \\
5 & 0x74bc...91fd6 & Absence of code logic or sanity check \\
6 & 0x0624...5286 & On-chain oracle manipulation \\
7 & 0x7b3b...6ca & Visibility errors, including unrestricted action \\
8 & 0x3212...8923 & Token standard incompatibility, Reentrancy \\
9 & 0x5bd6...dee4 & Unfair slippage protection, Absence of code logic or sanity check \\
10 & 0x32e5...de43 & Flash liquidity borrow, purchase, mint or deposit \\
11 & 0x39b1...db15 & On-chain oracle manipulation, Frontrunning \\
12 & 0x17e8...ca5f & Deployment mistake, Other unsafe DeFi protocol dependency \\
13 & 0x923c...d3a5f & Visibility errors, including unrestricted action \\
14 & 0xc1e0...24c5 & Governance attack \\
15 & 0x8809...4d61 & Token standard incompatibility \\
16 & 0x3ec4...8af0 & On-chain oracle manipulation \\
17 & 0x33bf...8a27 & Other fake contracts, Deployment mistake, Arithmetic mistakes \\
18 & 0x25a5...e5f0 & Absence of code logic or sanity check \\
19 & 0x951d...abfe2 & Absence of code logic or sanity check \\
20 & 0xe0b9...caed5 & Other coding mistakes \\
21 & 0x67b6...9c7a & Other unsafe DeFi protocol dependency, Absence of code logic or sanity check \\
22 & 0x9dae...2f9 & Other unsafe DeFi protocol dependency, Inconsistent access control \\
23 & 0x0e51...3682 & Other unsafe DeFi protocol dependency \\
24 & 0x7557...d8b & Visibility errors, including unrestricted action \\
25 & 0x6847...5210 & Absence of code logic or sanity check, Other unsafe DeFi protocol dependency \\
26 & 0x6684...64f & Other unsafe DeFi protocol dependency \\
27 & 0x35c6...7810 & Flash liquidity borrow, purchase, mint or deposit, Absence of code logic or sanity check \\
28 & 0x66e7...e63 & Absence of code logic or sanity check \\
29 & 0x55db...3830 & On-chain oracle manipulation, Absence of code logic or sanity check \\
30 & 0xddd7...1101 & On-chain oracle manipulation \\
31 & 0xc9f2...14ef & Reentrancy, Visibility errors, including unrestricted action, Delegatecall injection \\
32 & 0x0eee...5ea & Token standard incompatibility, Reentrancy \\
33 & 0x88cc...8b17 & Absence of code logic or sanity check \\
34 & 0xae46...8cf & On-chain oracle manipulation \\
35 & 0x818e...7755 & On-chain oracle manipulation \\
36 & 0xacbd...e747 & Flash liquidity borrow, purchase, mint or deposit \\
37 & 0x6b7a...522 & Unsafe call to phantom function, Absence of code logic or sanity check \\
38 & 0xa231...ef5 & Direct call to untrusted contract, Insider trade or other activities \\
39 & 0x929c...49d6 & Delegatecall injection, Absence of code logic or sanity check \\
40 & 0x3157...e758 & Governance attack \\
41 & 0x0e6f...a30 & Locked or frozen tokens \\
42 & 0x606e...962 & Absence of code logic or sanity check \\
43 & 0x1cec...2004 & Reentrancy \\
44 & 0x876b...dc8 & Visibility errors, including unrestricted action \\
45 & 0x328d...5424 & Reentrancy, Flash liquidity borrow, purchase, mint or deposit \\
46 & 0x77f9...81bc & Other unfair or unsafe DeFi protocol interaction, Absence of code logic or sanity check \\
47 & 0xde74...9b25 & Absence of code logic or sanity check \\
48 & 0xdac1...ec7 & Absence of code logic or sanity check \\
49 & 0xe7f4...58e6 & Absence of code logic or sanity check \\
50 & 0x0efb...c0b & Absence of code logic or sanity check, Fake tokens \\
51 & 0xe11f...f50 & Unfair slippage protection \\
52 & 0xacd4...952 & Flash liquidity borrow, purchase, mint or deposit, Unfair liquidity providing \\
\end{longtable}

\begin{longtable}{c l l c} % Use longtable instead of tabular
	\caption{Phase two results, smart contracts sourced from DeFi SoK [num]} \\
	\toprule
	ID & Address & Description & Success Rate (\%) \\
	\midrule
	\endfirsthead % Everything above this will be repeated on every page's header
	\caption{Phase two results (continued)} \\
	\toprule
	ID & Address & Description & Success Rate (\%) \\
	\midrule
	\endhead % Everything above this will be placed at the top of the continuation pages
	\bottomrule
	\endfoot % Footer for all pages except the last, usually empty
	\bottomrule
	\endlastfoot % Footer for the last page of the table
1 & 0x3ec4...8af0 & On-chain oracle manipulation & 80.0 \\
2 & 0x9059...2397 & Deployment mistake & 82.5 \\
3 & 0x9059...2397 & Absence of code logic or sanity check & 15.0 \\
4 & 0x0eee...f5ea & Token standard incompatibility & 60.0 \\
5 & 0x0eee...f5ea & Reentrancy & 27.5 \\
6 & 0x2775...9e38 & Absence of code logic or sanity check & 90.0 \\
7 & 0xde74...9b25 & Absence of code logic or sanity check & 0.0 \\
8 & 0x5a9f...6ed1 & Absence of code logic or sanity check & 30.0 \\
9 & 0x818e...b755 & On-chain oracle manipulation & 82.5 \\
10 & 0x27dd...4be8 & Absence of code logic or sanity check & 65.0 \\
11 & 0x929c...49d6 & Delegatecall / call injection & 20.0 \\
12 & 0x929c...49d6 & Absence of code logic or sanity check & 7.5 \\
13 & 0xe1e1...6c21 & Camouflage a token contract & 15.0 \\
14 & 0xe1e1...6c21 & Reentrancy & 97.5 \\
15 & 0x109e...a300 & On-chain oracle manipulation & 27.5 \\
16 & 0x304c...29ba & Compromised private key / hacked wallet & 30.0 \\
17 & 0x304c...29ba & Absence of code logic or sanity check & 72.5 \\
18 & 0xb89e...3258 & Backdoor / Honeypot & 85.0 \\
19 & 0x39b1...db15 & Frontrunning & 67.5 \\
20 & 0x39b1...db15 & On-chain oracle manipulation & 95.0 \\
21 & 0x8e96...6d27 & On-chain oracle manipulation & 37.5 \\
22 & 0x8e96...6d27 & Liquidity borrow, purchase, mint, deposit & 92.5 \\
23 & 0x8809...4d61 & Token standard incompatibility & 65.0 \\
24 & 0x9b3b...588b & On-chain oracle manipulation & 82.5 \\
25 & 0x88cc...8b17 & Known vulnerability not patched & 37.5 \\
26 & 0x3212...8923 & Token standard incompatibility & 90.0 \\
27 & 0x3212...8923 & Reentrancy & 62.5 \\
28 & 0xd7b7...0099 & Other protocol vulnerabilities & 95.0 \\
29 & 0xc145...c2ce & Absence of code logic or sanity check & 85.0 \\
30 & 0x25a5...e5f0 & Absence of code logic or sanity check & 85.0 \\
31 & 0x2165...b0a6 & Absence of code logic or sanity check & 90.0 \\
32 & 0x7b3b...c6ca & Function visibility error & 72.5 \\
33 & 0x8ae4...146d & Absence of code logic or sanity check & 75.0 \\
34 & 0x62cd...590f & Absence of code logic or sanity check & 32.5 \\
35 & 0x5377...ba2c & Direct call to untrusted contract & 85.0 \\
36 & 0x32e5...de43 & Liquidity borrow, purchase, mint, deposit & 67.5 \\
37 & 0xfb44...5df1 & Liquidity borrow, purchase, mint, deposit & 90.0 \\
38 & 0xfe67...3d1d & Liquidity borrow, purchase, mint, deposit & 70.0 \\
39 & 0x7ad8...9b39 & Liquidity borrow, purchase, mint, deposit & 67.5 \\
40 & 0xc178...ca62 & Backdoor / Honeypot & 67.5 \\
41 & 0x6684...a64f & Other unsafe DeFi protocol dependency & 35.0 \\
42 & 0xaf93...c0cc & Token standard incompatibility & 55.0 \\
43 & 0x54f2...5259 & Token standard incompatibility & 80.0 \\
44 & 0xe11f...df50 & Unfair slippage protection & 67.5 \\
45 & 0x5085...3369 & Other unsafe DeFi protocol dependency & 82.5 \\
46 & 0xe952...1659 & Token standard incompatibility & 67.5 \\
47 & 0xe952...1659 & Reentrancy & 0.0 \\
48 & 0xe200...ca82 & Deployment mistake & 67.5 \\
49 & 0xd8a0...5531 & Unintentional DoS & 90.0 \\
50 & 0xd8a0...5531 & Unintentional DoS & 87.5 \\
51 & 0x1027...e5db & Other inconsistent, improper or unprotected access control & 72.5 \\
52 & 0xa231...7ef5 & Insider trade or other activities & 87.5 \\
53 & 0xa231...7ef5 & Direct call to untrusted contract & 100.0 \\
54 & 0x17e8...ca5f & Deployment mistake & 65.0 \\
55 & 0x17e8...ca5f & Other unsafe DeFi protocol dependency & 85.0 \\
56 & 0x41a2...35c4 & On-chain oracle manipulation & 75.0 \\
57 & 0x41a2...35c4 & Unfair liquidity providing & 67.5 \\
58 & 0x6b56...03ab & Deployment mistake & 52.5 \\
59 & 0x45f8...674b & Deployment mistake & 82.5 \\
60 & 0x9930...7cc3 & Absence of code logic or sanity check & 67.5 \\
61 & 0x328d...5424 & Liquidity borrow, purchase, mint, deposit & 0.0 \\
62 & 0x328d...5424 & Reentrancy & 72.5 \\
63 & 0x06c2...355e & Absence of code logic or sanity check & 100.0 \\
64 & 0x833e...743d & On-chain oracle manipulation & 75.0 \\
65 & 0x99b0...252d & Other inconsistent, improper or unprotected access control & 67.5 \\
66 & 0xf17c...8ee2 & Other inconsistent, improper or unprotected access control & 30.0 \\
67 & 0xa386...421d & Camouflage a token contract & 20.0 \\
68 & 0xa386...421d & Reentrancy & 80.0 \\
69 & 0xa386...421d & Absence of code logic or sanity check & 82.5 \\
70 & 0xd613...e186 & Arithmetic mistakes & 20.0 \\
71 & 0x67db...fcf9 & On-chain oracle manipulation & 90.0 \\
72 & 0x00cc...9409 & On-chain oracle manipulation & 75.0 \\
73 & 0xe0b9...aed5 & Arithmetic mistakes & 92.5 \\
74 & 0x27d4...ea14 & On-chain oracle manipulation & 20.0 \\
75 & 0x27d4...ea14 & Liquidity borrow, purchase, mint, deposit & 55.0 \\
76 & 0x67b6...9c7a & Other unsafe DeFi protocol dependency & 25.0 \\
77 & 0x67b6...9c7a & Absence of code logic or sanity check & 85.0 \\
78 & 0xd5aa...edf2 & Other protocol vulnerabilities & 80.0 \\
79 & 0x0efb...1c0b & Camouflage a token contract & 42.5 \\
80 & 0x0efb...1c0b & Absence of code logic or sanity check & 30.0 \\
81 & 0x8f52...6088 & On-chain oracle manipulation & 40.0 \\
82 & 0x8f52...6088 & Unfair slippage protection & 37.5 \\
83 & 0x45f2...cdfa & Arithmetic mistakes & 10.0 \\
84 & 0x8407...f21e & Compromised private key / hacked wallet & 97.5 \\
85 & 0xd520...f834 & On-chain oracle manipulation & 77.5 \\
86 & 0x31b3...a2b2 & Other auxiliary vulnerabilities & 70.0 \\
87 & 0x37ee...5e6d & On-chain oracle manipulation & 82.5 \\
88 & 0x77f9...81bc & Other unfair or unsafe DeFi protocol interaction & 17.5 \\
89 & 0x77f9...81bc & Absence of code logic or sanity check & 85.0 \\
90 & 0xcd5f...6a3e & Backdoor / Honeypot & 75.0 \\
91 & 0x6847...5210 & Other unsafe DeFi protocol dependency & 80.0 \\
92 & 0x6847...5210 & Absence of code logic or sanity check & 77.5 \\
93 & 0xa32d...050a & Backdoor / Honeypot & 80.0 \\
94 & 0x46ba...b5f9 & Token standard incompatibility & 42.5 \\
95 & 0xb3e7...db4d & Token standard incompatibility & 82.5 \\
96 & 0xe774...a8ea & Absence of code logic or sanity check & 75.0 \\
97 & 0xddd7...1101 & On-chain oracle manipulation & 77.5 \\
98 & 0xc9f2...14ef & Reentrancy & 30.0 \\
99 & 0xc9f2...14ef & Delegatecall / call injection & 67.5 \\
100 & 0xc9f2...14ef & Function visibility error & 17.5 \\
101 & 0x52e4...029c & Backdoor / Honeypot & 82.5 \\
102 & 0x42a5...f27b & Absence of code logic or sanity check & 92.5 \\
103 & 0x956c...c151 & Deployment mistake & 87.5 \\
104 & 0xe7f4...58e6 & Absence of code logic or sanity check & 25.0 \\
105 & 0xc454...1f94 & Arithmetic mistakes & 60.0 \\
106 & 0x6b7a...1522 & Unsafe call to phantom function & 30.0 \\
107 & 0x6b7a...1522 & Absence of code logic or sanity check & 80.0 \\
108 & 0x2ec2...56cd & On-chain oracle manipulation & 75.0 \\
109 & 0xda20...fa75 & Absence of code logic or sanity check & 100.0 \\
110 & 0x7a8a...fc89 & Absence of code logic or sanity check & 100.0 \\
111 & 0x7a8a...fc89 & Other inconsistent, improper or unprotected access control & 72.5 \\
112 & 0x41b8...cf49 & Compromised private key / hacked wallet & 77.5 \\
113 & 0xd7cd...87b2 & Camouflage a token contract & 75.0 \\
114 & 0xd7cd...87b2 & Reentrancy & 92.5 \\
115 & 0xf1f8...99f0 & Absence of code logic or sanity check & 85.0 \\
116 & 0x274b...3700 & Liquidity borrow, purchase, mint, deposit & 100.0 \\
117 & 0x923c...3a5f & Function visibility error & 90.0 \\
118 & 0x8dfe...a9f7 & Function visibility error & 65.0 \\
119 & 0x5f58...667f & Function visibility error & 5.0 \\
120 & 0x4fe4...b4c4 & Backdoor / Honeypot & 30.0 \\
121 & 0x374a...9233 & Other coding mistakes & 82.5 \\
122 & 0x4e3f...9d2b & Governance attack & 100.0 \\
123 & 0x1bba...304d & Absence of code logic or sanity check & 95.0 \\
124 & 0xacd4...f952 & Liquidity borrow, purchase, mint, deposit & 50.0 \\
125 & 0xacd4...f952 & Unfair liquidity providing & 67.5 \\
126 & 0xc145...c2ce & Absence of code logic or sanity check & 85.0 \\
127 & 0xbdad...8d43 & Reentrancy & 32.5 \\
128 & 0xd63b...3546 & Absence of code logic or sanity check & 77.5 \\
129 & 0x0e51...3682 & Other unsafe DeFi protocol dependency & 85.0 \\
130 & 0x6b7a...1522 & Other coding mistakes & 77.5 \\
131 & 0xd1c5...b5f3 & Other coding mistakes & 72.5 \\
132 & 0x88e6...5640 & On-chain oracle manipulation & 70.0 \\
133 & 0x0b28...d7ca & Backdoor / Honeypot & 90.0 \\
134 & 0x2688...b9fb & Camouflage a token contract & 42.5 \\
135 & 0x2688...b9fb & Liquidity borrow, purchase, mint, deposit & 27.5 \\
136 & 0xcaf0...5f91 & Unfair liquidity providing & 77.5 \\
137 & 0xa08c...9a4a & Broken patch & 32.5 \\
138 & 0xdd0c...886a & Broken patch & 65.0 \\
139 & 0x0fbb...4966 & Broken patch & 75.0 \\
140 & 0x0453...2148 & Broken patch & 7.5 \\
141 & 0x9b9b...caff & Broken patch & 7.5 \\
142 & 0x55db...3830 & On-chain oracle manipulation & 100.0 \\
143 & 0x55db...3830 & Absence of code logic or sanity check & 80.0 \\
144 & 0x4071...370c & Direct call to untrusted contract & 97.5 \\
145 & 0xae46...c8cf & On-chain oracle manipulation & 67.5 \\
146 & 0x45f7...5f51 & On-chain oracle manipulation & 7.5 \\
147 & 0x951d...bfe2 & Absence of code logic or sanity check & 80.0 \\
148 & 0x250e...c906 & Absence of code logic or sanity check & 100.0 \\
149 & 0x2f7a...1f03 & Absence of code logic or sanity check & 92.5 \\
150 & 0x8927...1d07 & On-chain oracle manipulation & 40.0 \\
151 & 0xa79f...759a & On-chain oracle manipulation & 67.5 \\
152 & 0x584d...6a6d & Token standard incompatibility & 92.5 \\
153 & 0x4b38...5b92 & Token standard incompatibility & 52.5 \\
154 & 0xb246...42cd & Backdoor / Honeypot & 72.5 \\
155 & 0x1ebd...ea20 & Authority control or breach of promise & 7.5 \\
156 & 0xe777...b02d & Authority control or breach of promise & 40.0 \\
157 & 0xa416...bdf6 & Authority control or breach of promise & 72.5 \\
158 & 0xcc25...e13e & Authority control or breach of promise & 27.5 \\
159 & 0xd361...11e1 & Authority control or breach of promise & 27.5 \\
160 & 0x450a...1927 & Authority control or breach of promise & 10.0 \\
161 & 0x7755...4dab & Authority control or breach of promise & 2.5 \\
162 & 0x8cc2...0a19 & Authority control or breach of promise & 2.5 \\
163 & 0x719e...1fc8 & Authority control or breach of promise & 60.0 \\
164 & 0x8770...fb90 & Authority control or breach of promise & 5.0 \\
165 & 0xefcb...5b8f & Authority control or breach of promise & 70.0 \\
166 & 0x71e1...d696 & Authority control or breach of promise & 60.0 \\
167 & 0x54d6...a0d9 & Authority control or breach of promise & 30.0 \\
168 & 0x8cc7...214e & Authority control or breach of promise & 30.0 \\
169 & 0x4265...eab6 & Authority control or breach of promise & 82.5 \\
170 & 0xf3d1...0e12 & Authority control or breach of promise & 100.0 \\
171 & 0x748f...190d & Authority control or breach of promise & 77.5 \\
172 & 0x819e...9b5d & On-chain oracle manipulation & 97.5 \\
173 & 0x0624...5286 & On-chain oracle manipulation & 72.5 \\
174 & 0x65bc...054f & On-chain oracle manipulation & 30.0 \\
175 & 0x74bc...1fd6 & Absence of code logic or sanity check & 30.0 \\
176 & 0x1bba...304d & On-chain oracle manipulation & 50.0 \\
177 & 0x66e7...ee63 & Absence of code logic or sanity check & 67.5 \\
178 & 0x7087...085b & Absence of code logic or sanity check & 65.0 \\
179 & 0xef9c...cc6b & Compromised private key / hacked wallet & 35.0 \\
180 & 0x606e...8962 & Absence of code logic or sanity check & 10.0 \\
181 & 0xdac1...1ec7 & Broken patch & 27.5 \\
182 & 0x5bd6...dee4 & Unfair slippage protection & 2.5 \\
183 & 0x5bd6...dee4 & Absence of code logic or sanity check & 82.5 \\
184 & 0xc1e0...24c5 & Governance attack & 25.0 \\
185 & 0x328d...5424 & Liquidity borrow, purchase, mint, deposit & 72.5 \\
186 & 0x328d...5424 & Reentrancy & 17.5 \\
187 & 0xa2a2...f2b1 & Absence of code logic or sanity check & 92.5 \\
188 & 0xfd55...1001 & Absence of code logic or sanity check & 30.0 \\
189 & 0x33bf...8a27 & Deployment mistake & 92.5 \\
190 & 0x33bf...8a27 & Camouflage a non-token contract & 82.5 \\
191 & 0x33bf...8a27 & Arithmetic mistakes & 72.5 \\
192 & 0x876b...1dc8 & Function visibility error & 72.5 \\
193 & 0x2bbd...1445 & Function visibility error & 72.5 \\
194 & 0xacbd...e747 & Liquidity borrow, purchase, mint, deposit & 40.0 \\
195 & 0x55db...3830 & On-chain oracle manipulation & 47.5 \\
196 & 0x55db...3830 & Absence of code logic or sanity check & 70.0 \\
197 & 0x0e6f...4a30 & Improper asset locks or frozen asset & 57.5 \\
198 & 0x7ac5...e677 & Camouflage a token contract & 35.0 \\
199 & 0x7ac5...e677 & Reentrancy & 10.0 \\
200 & 0x7ac5...e677 & Absence of code logic or sanity check & 35.0 \\
201 & 0xcb6a...a723 & Other protocol vulnerabilities & 27.5 \\
202 & 0xcb6a...a723 & Absence of code logic or sanity check & 57.5 \\
203 & 0x1cec...2004 & Reentrancy & 40.0 \\
204 & 0xe4fe...9410 & Camouflage a token contract & 37.5 \\
205 & 0xe4fe...9410 & Absence of code logic or sanity check & 72.5 \\
206 & 0xd065...60ee & Token standard incompatibility & 10.0 \\
207 & 0xd065...60ee & Reentrancy & 67.5 \\
208 & 0x2db6...66d6 & Token standard incompatibility & 85.0 \\
209 & 0x2db6...66d6 & Reentrancy & 77.5 \\
210 & 0x7557...1d8b & Function visibility error & 82.5 \\
211 & 0x35c6...7810 & Liquidity borrow, purchase, mint, deposit & 75.0 \\
212 & 0x35c6...7810 & Absence of code logic or sanity check & 95.0 \\
213 & 0x3157...e758 & Governance attack & 97.5 \\
214 & 0x14e6...e4bd & Frontrunning & 12.5 \\
215 & 0x14e6...e4bd & Direct call to untrusted contract & 72.5 \\
216 & 0xa800...190d & Unfair liquidity providing & 65.0 \\
217 & 0xd55f...dba5 & On-chain oracle manipulation & 5.0 \\
218 & 0x9dae...f2f9 & Other unsafe DeFi protocol dependency & 82.5 \\
219 & 0x9dae...f2f9 & Other inconsistent, improper or unprotected access control & 85.0 \\
\end{longtable}

\end{document}